\documentclass[12pt]{article}

\usepackage[totalheight = 23cm, totalwidth = 17cm]{geometry}
\usepackage{amssymb,amsmath,amsfonts,amsbsy,graphicx,bm}

\usepackage{color,cancel,ulem, hyperref}

\newcommand{\dis}[1]{\begin{equation}\begin{split}#1\end{split}\end{equation}}

\begin{document}

\begin{titlepage}

\begin{center}

{\setlength{\baselineskip}%
{1.5\baselineskip}
{\LARGE \bf 
Implication of dressed form of relational observable  on  von Neumann algebra
}
\par}

\vskip 1.0cm

{\large
Min-Seok Seo$^{a}$ 
\footnote[0]{e-mail : minseokseo57@gmail.com}
}

\vskip 0.5cm

{\it
$^{a}$Department of Physics Education, Korea National University of Education,
\\ 
Cheongju 28173, Republic of Korea
}

\vskip 1.2cm

\end{center}

\begin{abstract}

  In quantum gravity, a physically meaningful operator is required to be invariant under the diffeomorphisms.
  Such gauge invariant operator is typically given by the relational observable, the   operator localized  in relation to some background states.
  We point out that the relational observable can be comprehensively formulated as  the dressed operator.
  For the background with a boundary where the diffeomorphisms are not gauged, we can use the gravitational Wilson line for dressing, then the relational observable is nonlocal.
  In contrast, when the background breaks some isometries, as can be found in quasi-de Sitter space, dressing can be local, implementing St\"uckelberg mechanism.
   Since dressing resembles the outer automorphism in the von Neumann algebra, we may investigate the algebraic structure of the background by considering the dressed form of the relational observable.
   From this, we can understand that quasi-de Sitter space is described by the Type II$_\infty$ algebra where the trace diverges in the decoupling limit of gravity.
   It is different from the Type II$_1$ algebra of de Sitter space where the finite size  of trace can be defined in the same limit.
   This shows that the isometry preserving and breaking backgrounds are quite different in the algebraic structure   even in the decoupling limit of gravity.

\end{abstract}

\end{titlepage}

\newpage

\section{Introduction}

 General relativity is now widely accepted as the standard  effective field theory (EFT) description of gravity at low energy scale.
 In this framework, spacetime is a dynamical field, which is formulated by treating diffeomorphisms as gauge symmetries.
 Then  we expect  that  a physically meaningful operator   in quantum gravity would be invariant under the diffeomorphisms. 
 Such gauge invariant operator typically takes the form of  the `relational observable', which is obtained by localizing the operator  in  relation to some background states playing the role of the clock and the rod \cite{Rovelli:1990jm, Rovelli:1990ph, Rovelli:1990pi, Rovelli:1989jn} (see \cite{Komar:1958ymq, DeWitt:1962cg, DeWitt:1967yk} for earlier discussions and also \cite{Marolf:1994wh, Marolf:1994nz, Giddings:2005id, Dittrich:2004cb, Dittrich:2005kc, Tambornino:2011vg, Goeller:2022rsx, FrancoisAndre:2023jmj, Francois:2024rdm, Francois:2025shu, Francois:2025lqn}).

  The first step toward constructing the relational observable is to relate every point in spacetime to the clock and the rod through functionals of the metric and the matter fields (see \cite{Giddings:2025xym} for a recent review).
   If the functionals depend on the metric only, the relational observable is said to be `gravitationally dressed' \cite{Donnelly:2015hta, Donnelly:2016rvo, Giddings:2018umg, Giddings:2022hba}.  
 It can be easily constructed when the background is asymptotically Minkowski or anti de Sitter (AdS), which can be continued to the   `boundary', namely, an asymptotic infinity or the AdS boundary, where gravity is not taken to be dynamical,  hence diffeomorphisms are not gauged.
  This enables one to implement the region called the `platform' at the boundary,  where points are also  fixed under the diffeomorphisms.
  Then the relational observable is constructed by dressing  the local operator by the gravitational Wilson line along the geodesic (hence given by the functional of the metric) connecting  a  spacetime point to a point on the platform.
  A notable feature of this procedure is that, while the dressed operator becomes gauge invariant, it is no longer local.
  Indeed, attempts in \cite{Geng:2020fxl, Geng:2020qvw, Geng:2021hlu, Geng:2023zhq, Geng:2024dbl, Geng:2025bcb, Geng:2025gqu} claimed that the locality of operators is maintained if we allow the energy to flow into the boundary.
   In this case,  diffeomorphisms are broken by the boundary conditions, resulting in the massive graviton.
  
   On the other hand, it has been pointed out that the tension between gauge invariance and locality arises from isometries of the background \cite{Antonini:2025sur} (see also \cite{Seo:2025tsw}).
   When the background slightly breaks the isometry, the classical solutions to the matter field as well as the background distinguish different slices of the coordinates in the direction of broken isometry.
    Then they can be used as the clock and the rod, from which we can implement the functionals of  the matter or the metric fields to construct the local relational observable.
    As we will see, this procedure can be interpreted as the St\"uckelberg mechanism: in order to form the local gauge invariant operator, some of the metric fluctuations combine  with the matter fluctuation.    
   Since the transverse  traceless fluctuation of the metric is gauge invariant by itself, it does not contribute to the St\"uckelberg mechanism.
   Therefore, the graviton remains massless and diffeomorphisms are not broken.

   Such local relational observables have been extensively studied in the context of cosmology to describe the cosmological fluctuations, where the background does not have a well-defined boundary,  hence the platform    \cite{Bardeen:1980kt, Sasaki:1983kd, Brandenberger:1983vj, Halliwell:1984eu, Shirai:1987xh, Mukhanov:1988jd, Deruelle:1991sb}.
   In particular, the curvature perturbation, the scalar fluctuation during the inflationary era, is given by a combination of the fluctuation of the trace part of the metric and that of the inflaton, hence can be interpreted as a local relational observable resulting from the breaking of timelike de Sitter (dS) isometry  by the  quasi-dS background  \cite{Mukhanov:1985rz, Sasaki:1986hm, Cheung:2007st, Weinberg:2008hq} (see also \cite{Prokopec:2010be, Gong:2016qpq}). 
   It provides the standard explanation for the primordial inhomogeneities appearing in the cosmic microwave background  \cite{Mukhanov:1981xt, Mukhanov:1990me}.

   The purpose of this article is to understand the features of the local relational observable arising from the  broken isometry and the nonlocal gravitationally dressed operator in a comprehensive way : as discussed in section \ref{Sec:Str}, both can be formulated in the form of the dressed local field operator.
   The difference arises from whether the dressing is local or not.
   Intriguingly, the form of  dressed operator resembles the outer automorphism of the local operator appearing in the von Neumann algebraic description of quantum field theory in curved spacetime \cite{Giddings:2022hba, Geng:2025bcb} (for recent reviews on von Neumann algebra, see, e.g., \cite{Witten:2018zxz, Witten:2021jzq, Sorce:2023fdx, Liu:2025krl}).
   Indeed, the von Neumann algebra of the isometry breaking background seems to be drastically different from that of the background with isometry (see also \cite{Grassi:2024vkb, deSabbata:2026aum} for studies in the similar direction to ours).  
  We can see this by comparing quasi-dS space   with perfect dS space.
  For dS space, the absence of the clock and the rod in the background compels us to introduce an observer carrying the clock and the rod.
 Imposing that the energy of an observer is positive, one can define a trace of an operator of   finite size, from which we can consider various thermodynamic quantities even in the limit $\kappa=(8\pi G)^{1/2} \to 0$  \cite{Chandrasekaran:2022cip} (see also \cite{Aguilar-Gutierrez:2023odp}).
 This corresponds to the Type II$_1$ von Neumann algebra.
 In contrast, for quasi-dS space where the timelike isometry is broken, the background or the inflaton field provides a clock.
 Then, as is also observed in the case of black hole, the fluctuation of the renormalized energy diverges to $\pm \infty$  in the $\kappa\to 0$ limit, which originates from the accumulation of the scalar field fluctuations stretched outside the horizon (see a brief review in Appendix \ref{App:qdScase}).
 This is reflected in the divergence of trace in the same limit,   a characteristic of the Type II$_\infty$ von Neumann algebra \cite{Seo:2022pqj, Gomez:2023wrq, Kudler-Flam:2024psh, Chen:2024rpx}.
 In  Section \ref{Sec:vNtype} we investigate the implications of the relational observable in the dressed form on the von Neumann algebra  in detail.
   Indeed, whereas \cite{Seo:2022pqj, Gomez:2023wrq, Kudler-Flam:2024psh} showed how the scalar field fluctuations stretched outside the horizon leads to Type II$_\infty$ algebra, they introduced an observer.
   While an observer may be required to describe the entire evolution of the universe (as considered in \cite{Kudler-Flam:2024psh}), so fas as we restrict our attention to the inflationary period when the timelike dS isometry is slightly broken, we do not need to introduce an observer playing the role of clock.
  Then we only need to introduce an observer as a rod, which is essential to define the local subregion since the spatial homogeneity and isotropy are still maintained.  
 This fact   was pointed out in \cite{Chen:2024rpx}, but  a rolling scalar on the tilted  potential is simply considered on the perfect dS space which requires the flat potential.
 While this is a good approximation since the timelike dS isometry breaking is typically tiny, the connection between the role of the scalar field as a clock and the isometry breaking by the quasi-dS background can be overlooked.
 As a result, an important feature of quasi-dS space resulting from breaking of the timelike dS isometry may be missed: in quasi-dS space, the fluctuation of the scalar field is connected to that of time,  hence to that of metric, through the St\"uckelberg mechanism.
 In this article, we try to clarify how such a feature leads to algebraic structure of quasi-dS space drastically different from that of perfect dS space, emphasizing the role of gauge invariance which reflects the dynamic nature of gravity.

\section{Relational observable as dressed operator}  
\label{Sec:Str}
  
  In this section, we discuss the common structure of the relational observable which can be found in both the local relational observable and the nonlocal gravitationally dressed one.
  To see this, let us recall the primary motivation of the  relational observable,   the difficulty in finding the local  gauge invariant operator: even a scalar field operator $O(p)$ ($p$ :  a point in the manifold ${\cal M}$)   is not invariant under the active diffeomorphism $p \mapsto \phi (p)$ as it transforms as $O(p) \mapsto O'(p)=O(\phi^{-1}(p))$. 
 For an infinitesimal diffeomorphism $x \mapsto x+\kappa\xi$, where $x$ denotes the coordinates of $p$ we are working on,  the transformed scalar field is given by $O'(x)=O(x-\kappa\xi)$, corresponding to the field variation $\delta O(x)=-\kappa\xi^\mu \partial_\mu O(x)$.
 The relational observable resolves this gauge invariance issue by depending on the point specified by the values of the clock and the rod operators given by, for example, scalar operators $X^a (x)$ ($a=1, \cdots, {\rm dim}({\cal M})=4$).
 More concretely, imposing the condition 
\dis{X^a(x) =y^a,\label{eq:ConRel}} 
where the values of $y^a$ are fixed under the diffeomorphism, the relational observable depends on $x^\mu$ through its inverse function $x^\mu=\chi^\mu(y^a)$ as  $O(\chi(y^a))$.
  Since $X^a(p) \mapsto {X'}^a(p)=X^a(\phi^{-1}(p))$  under the diffeomorphism,  the condition \eqref{eq:ConRel} is satisfied by $p$ and $\phi(p)$ before and after the transformation, respectively, i.e., $X^a(p)= {X'}^a(\phi(p))=y^a$.
  Then the transformed relational observable becomes $O'(\phi(p))$, rather than $O'(p)$, from which  gauge invariance is achieved as $O'(\phi(p))=O(\phi^{-1}(\phi(p)))=O(p)$.   
 When the diffeomorphism is infinitesimal, it reads
 \dis{O(x)\mapsto O'(x+\kappa\xi)&=O'(x)+\kappa\xi^\mu \partial_\mu O'(x) 
 \\
 &=O(x)-\kappa\xi^\mu \partial_\mu O(x)+\kappa\xi^\mu \partial_\mu O(x) =O(x).}
 
 Since the construction of the relational observable requires an additional transformation of $p$, it is tempting to express it in the `dressed form'.
 For the local scalar operator $O_M(x)$ consisting of the matter fields only, it is written as   
\dis{O_{\rm dr}=e^{-i H_M[q]}O_M(x) e^{iH_M[q]},\label{eq:DrO}}
where 
\dis{H_M[q] =-\int_\Sigma  d\Sigma  T_{\mu\nu} q^\mu n^\nu=-\int_\Sigma {{\cal H}_M}_\mu q^\mu} 
is the  diffeomorphism  generator translating $x$ by $q$ ($\Sigma$ :   Cauchy surface,   $n^\nu$ :  timelike unit normal to $\Sigma$).
Identifying $n^\nu$ with the timelike normal vector in the Arnowitt-Deser-Misner (ADM) foliation, one obtains the last equality where  ${{\cal H}_M}_\mu$ denotes the matter part of the ADM Hamiltonian ($\mu=0$) and momentum densities ($\mu=1,2,3$) \cite{Arnowitt:1962hi}.
It is evident from $\delta O_M(x)=-i[O_M(x), H_M[\kappa\xi]]=-\kappa \xi^\mu \partial_\mu O_M(x)$ that the dressing to define $O_{\rm dr}$ corresponds to the shift $x \mapsto x+q$ of $x$ on which $O_M(x)$ depends.
For \eqref{eq:DrO} to be gauge invariant, $q^\mu$ is given by the functional of the metric fluctuation and transforms in the same way as $x^\mu$ under the diffeomorphism.
That is, under the infinitesimal diffeomorphism, it shifts as 
\dis{\delta q^\mu=-i[q^\mu, H_G[\kappa\xi]]=\kappa \xi^\mu \label{eq:qcond}} where 
\dis{H_G[\kappa\xi]=-\int_\Sigma {{\cal H}_G}_\mu \kappa \xi^\mu, }
with   ${{\cal H}_G}_\mu$ being the gravity part of the ADM Hamiltonian and momentum densities. 
Then $F[q]$, the functional of $q$, varies as  $\delta F[q]=-i[F[q], H_G[\kappa\xi]]=(\delta F/\delta q^\mu)[q]\cdot\kappa\xi^\mu$.
Noting that $[H_M[\kappa\xi], H_G[\kappa\xi]]=0=[q^\mu, H_M[\kappa\xi]]$, one finds 
 \dis{\delta O_{\rm dr}&=-i[O_{\rm dr}, H_G[\kappa\xi]+H_M[\kappa\xi]]
 \\
 &=\big(-i[e^{-i H_M[q]}, H_G[\kappa\xi]]\big)O_M(x)e^{i H_M[q]}
 +e^{-i H_M[q]}O_M(x)\big(-i[e^{i H_M[q]}, H_G[\kappa\xi]]\big)
 \\
 &\quad\quad +e^{-i H_M[q]}\big(-i[O_M(x), H_M[\kappa\xi]]\big) e^{i H_M[q]}
 \\
 &=e^{-i H_M[q]}\big(i[O_M(x), H_M[\kappa\xi]]\big) e^{i H_M[q]}+e^{-i H_M[q]}\big(-i[O_M(x), H_M[\kappa\xi]]\big) e^{i H_M[q]} =0,}
 showing that $O_{\rm dr}$ is gauge invariant. 
 
 We now discuss the nonlocal gravitationally dressed operator.
 To construct the functional $q$ of the metric fluctuation satisfying \eqref{eq:qcond}, it uses the fact that the geodesic equation is covariant under the diffeomorphism \cite{Donnelly:2015hta}.  
 For the asymptotically Minkowski or AdS background, the asymptotic infinity or the AdS boundary can be regarded as the `boundary' where gravity may not be taken to be dynamical, hence diffeomorphism is not gauged.
  Then we can implement the 3-dimensional (more precisely, $({\rm dim}{\cal M}-1)-$dimensional) subregion  called  `platform'  located at infinitely large fixed value of, say, $x^1$  (hence it belongs to the boundary region). 
 With this setup, any point in spacetime can be specified by the geodesic $V^\mu$ launched perpendicular to the platform, which is a functional of the metric fluctuation determined by the coordinates $(y^0, 0, y^2, y^3)$ on the platform and the proper distance $s$ from the platform to the point.
  That is, the position of the point is written as $x^\mu=y^\mu+V^\mu$, and since $y^\mu$ and $s$ are fixed under the diffeomorphism (hence correspond  to $y^a$ in \eqref{eq:ConRel}), the covariance of $x^\mu$ under the diffeomorphism is achieved by $\delta V^\mu=-i[V^\mu, H_G[\kappa\xi]]=\kappa \xi^\mu$.
 For example, if the spacetime geometry is given by the small fluctuation of the flat spacetime, the metric can be written as $g_{\mu\nu}=\eta_{\mu\nu}+\kappa h_{\mu\nu}$, and the geodesic equation for $u^\mu=dV^\mu/ds$,
 \dis{\frac{d u^\mu}{ds}=-\Gamma^\mu_{~\nu\lambda}u^\nu u^\lambda=-\frac{\kappa}{2}\Big(\partial_\nu h^\mu_{~\lambda}+\partial_\lambda h^\mu_{~\nu}-\partial^\mu h_{\nu\lambda}\Big)u^\nu u^\lambda+{\cal O}(\kappa^2),}
 is solved to give

 \dis{x_\mu&=y_\mu+\frac{\kappa}{2}\int_y^\infty d{x'}^\nu\Big[h_{\mu\nu}(x')+\int_{x'}^\infty d{x''}^\lambda \big(\partial_\mu h_{\nu\lambda}(x'')-\partial_\nu h_{\mu\lambda}(x'')\big)\Big]
 \\
 &=y_\mu+V_\mu.}
Since $\delta h_{\mu\nu}=-\partial_\mu \xi_\nu-\partial_\nu\xi_\mu$, the infinitesimal diffeomorphism transforms $V^\mu$ into $V^\mu+\kappa \xi^\mu$, as expected \cite{Donnelly:2015hta}.
 For more generic background, we refer the reader to \cite{Giddings:2022hba}.
 
 In summary, the nonlocal gravitationally dressed operator written in the form of 
 \dis{O_{\rm dr}=e^{-i H_M[V]}O_M(x)e^{i H_M[V]}\label{eq:Drflat}}
 can be interpreted that  $x$ where the operator $O_M$ is defined is translated to a point in the platform $y^\mu=x^\mu -V^\mu$ by the proper distance $s$.
 The fact that  both $y^\mu$ and $s$  do not transform under the diffeomorphism guarantees   gauge invariance of  $O_{\rm dr}$.
 Since the metric fluctuation contributes to the geodesic $V^\mu$ through the connection $\Gamma^\mu_{~\nu\lambda}$,   $O_{\rm dr}$ is said to be dressed by the gravitational Wilson line connecting the point to the platform.
It is notable that the construction of $O_{\rm dr}$ in this way accomplishes   gauge invariance at the price of promoting the local operator $O_M(x)$ to the nonlocal one.
 
 On the other hand, when the background breaks the isometry slightly, we can construct the relational observable without violating the locality.
 Even in this case, the local relational observable can be written in the dressed form  given by \eqref{eq:DrO}.
To see this, we consider the inflationary cosmology, during which the quasi-dS background slightly deviates from dS  space such that the timelike isometry is broken.
 Two backgrounds, dS and quasi-dS spaces, can be easily compared  in the flat coordinates, where the metric   is given by 
 \dis{ds^2=-dt^2+a(t)^2\big(dr^2+r^2(d\theta^2+\sin^2\theta d\varphi^2)\big),\label{eq:flatMet}} 
from which we   define the   Hubble parameter $H=\dot{a}/a$ (here dot denotes the derivative with respect to $t$).
For dS space, $H$ is constant, hence $a(t)=e^{Ht}$, which means that the event horizon coincides with the apparent horizon and the radius is given by  $R_{\rm hor}=1/H$.
 Moreover,  the translation of $t$ combines with the scaling of $r$ to become the timelike dS isometry.
 This is evident in the static coordinates,
 \dis{t_s=t-\frac{1}{2H}\log\big(1-H^2r^2e^{2Ht}\big),\quad r_s=r e^{Ht},\label{eq:conv}}
where the components of the metric
 \dis{ds^2=-(1-H^2r_s^2)dt^2+\frac{dr_s^2}{1-H^2r_s^2}+r_s^2(d\theta^2+\sin^2\theta d\varphi^2)}
 is independent of the static time $t_s$.
 Then the timelike Killing vector is given by $k^a=(\partial_{t_s})^a=(\partial_t-Hr\partial_r)^a$.
 In contrast, for quasi-dS space, $H$ slowly varies in time, distinguishing time slices specified by the values of $t$.
 Then the timelike isometry of dS space is no longer respected, which can be parametrized by the slow-roll parameter defined by 
 \dis{\epsilon_H=\dot{R}_{\rm hor}=-\frac{\dot H}{H^2}=\frac{\kappa^2}{2}\frac{\dot{\phi}_0^2}{H^2}.\label{eq:epsilon}} 
The last equality shows that when the timelike isometry is broken, the classical solution to the inflaton field $\phi_0$ is no longer constant, but depends on $t$.
 In other words, not only the value of $H$, but also that of  $\phi_0$   can be used as the clock specifying the time slice of constant $t$.

  A remarkable feature of quasi-dS space is that the fluctuation of the trace part of the metric which  was not physical in the presence of the timelike isometry combines with the inflaton fluctuation to become physical.
  This can be found from the infinitesimal transformations of the metric and the inflaton fluctuations   in the direction of $t$,
  \footnote{Here the fluctuation of the metric $h_{ij}(t, \mathbf{x})$ and that of the inflaton $\varphi(t, \mathbf{x})$ are defined by
  \dis{\gamma_{ij}(t, \mathbf{x})=a(t)^2\big[\delta_{ij}+\kappa h_{ij}(t, \mathbf{x})\big],\quad\quad \phi(t, \mathbf{x})=\phi_0(t)+\varphi(t, \mathbf{x})}
  in the ADM decomposition $ds^2=-N^2 dt^2+\gamma_{ij}(N^i+dx^i)(N^jdt+dx^j)$ with $N=1$ and $N^i=0$. }
  \dis{&\delta h_{ij}=-i\Big[h_{ij},   H_G[\kappa\epsilon] \Big]=-2H\epsilon \delta_{ij} ,
  \\
  &\delta \varphi=-i\Big[\varphi, H_M [\kappa\epsilon]\Big]=-\dot{\phi}_0 \kappa\epsilon,\label{eq:shift}}
  where $H_{G/M}$  generates  the translation in the direction of $t$  and is given by $-\int_\Sigma {\cal H}^0_{G/M}\kappa\epsilon$ (${\cal H}^0_{G/M}$ : the gravity/matter part of the Hamiltonian).
  Then both the fluctuation of the metric trace $\zeta$ defined by $h_{ij}=2\zeta\delta_{ij}+\cdots$ (hence $\delta \zeta=-H\epsilon$) and that of the inflaton shift under the infinitesimal time translation.
  From this, one immediately finds the gauge invariant combination called the Mukhanov-Sasaki variable \cite{Mukhanov:1985rz, Sasaki:1986hm},
  \dis{{\cal R}=\zeta-\frac{H}{\kappa\dot{\phi}_0}\varphi=\zeta-\frac{1}{\sqrt{2\epsilon_H}}\varphi,\label{eq:MS}}
  which is interpreted as being constructed by the St\"uckelberg mechanism (or the Higgs mechanism in a broad sense): just as the unphysical longitudinal mode of the gauge boson becomes physical by absorbing the pseudoscalar, the unphysical metric fluctuation $\zeta$ absorbs $\varphi$ to become a physical, gauge invariant operator.

  Since $\zeta$ shifts under the   infinitesimal $t$ translation, it can play a role of $q$ in \eqref{eq:DrO}.
  Noting that $-(\kappa/H)\delta\zeta=\kappa \epsilon$ under the infinitesimal time translation, we can construct the time translation invariant operator in the dressed form,
\dis{O_{\rm dr}[\varphi]=e^{-i H_M[-\frac{\kappa}{H}\zeta]}O_M(x) e^{iH_M[-\frac{\kappa}{H}\zeta]},\label{eq:DrIn}}
or equivalently, $O_M(t -\frac{\kappa}{H}\zeta, \mathbf{x})$, where $O_M(x)$ is made up of the matter (inflaton in this case) field.
When the operator $O_M$ is the inflaton field $\phi$ itself, it can be expanded around $\phi_0(t)$ as
\dis{\phi\Big(t-\frac{\kappa}{H}\zeta, \mathbf{x}\Big)=\phi_0(t)-\frac{\kappa}{H}\zeta \dot{\phi}_0 +\varphi +{\cal O}(\kappa^2, \varphi^2)=\phi_0(t)-\frac{\kappa \dot{\phi}_0}{H}{\cal R}(t, \mathbf{x}),}
showing that the Mukhanov-Sasaki variable is interpreted as a promotion of the inflaton field to the dressed one. 
 Moreover, since $\varphi$ also shifts under  the $t$ translation,  we can promote the operator $O_G(x)$ made up of the metric fluctuation $h_{\mu\nu}$ to   the $t$ translation  invariant one, which indeed is in the form of the dressed operator
  \footnote{While  the metric fluctuations relevant to the low energy EFT  are $\zeta$ and the graviton (transverse traceless mode), the latter is gauge invariant by itself (see Box 35.1 of \cite{Misner:1973prb}),  so we just need to promote the operator made up of $\zeta$ to the gauge invariant one by the dressing.}  
  :
  \dis{O_{\rm dr}[h]=e^{-iH_G[-\frac{\varphi}{\dot{\phi}_0}]}O_G(x) e^{ iH_G[-\frac{\varphi}{\dot{\phi}_0}]}.\label{eq:Drh}}

It is notable that whereas both \eqref{eq:Drflat} and \eqref{eq:DrIn} (and also \eqref{eq:Drh})  have the same form as  \eqref{eq:DrO} hence can be interpreted as the dressed operator realizing  the relational observable, they differ in the way of realization.
Dressing in  \eqref{eq:Drflat}  localizes the operator using the gravitational Wilson line  connecting a point in spacetime to another point on the platform.
The point in the platform, together with the proper distance,  plays the role of the clock and the rod since the platform is located in the boundary where the points are fixed under the diffeomorphism.
Since the gravitational Wilson line is a nonlocal operator,  the dressed operator  achieves gauge invariance in a nonlocal way.
In contrast, when the isometry is  broken by the background, the gauge invariant dressed operator is local.
Taking the quasi-dS background  as an example, the background (and also $\phi_0(t)$)  breaks the timelike dS isometry, distinguishing different time slices.
In other words, the background itself plays the role of the clock so we do not need to introduce the additional platform.  
Moreover, under the diffeomorphism some local field fluctuations  ($\zeta$ and $\varphi$ in this case) shift  in the direction of the broken isometry, thus can be used in the dressing.
As a result,   gauge invariance is achieved by the local dressed operator.
We also note that in the case of the  broken isometry, the metric fluctuation used in the dressing is not the transverse  traceless mode, showing that the graviton remains massless and diffeomorphism is not broken.
 For details including the broken spacelike isometry, we refer the reader to \cite{Seo:2025tsw}.

 \section{Implication of dressing on von Neumann algebra}
 \label{Sec:vNtype}
 
We now discuss how the structure of the relational observable can be related to  the von Neumann algebra, which has been investigated to understand the operator algebra and the thermodynamic behavior of quantum gravity comprehensively (see \cite{Witten:2018zxz, Witten:2021jzq, Sorce:2023fdx, Liu:2025krl} and references therein).
There are several reasons to address this issue.
First, in the  von Neumann algebra, a local operator  is promoted by the outer automorphism to a form identical to  the dressed operator  given by \eqref{eq:DrO}, achieving gauge invariance   \cite{Giddings:2022hba, Geng:2025bcb, Giddings:2025xym}. 
Second, it has been argued that whereas the von Neumann algebra for dS space is of Type II$_1$ (where the trace of operator converges hence is well defined even in the $\kappa \to 0$ limit) \cite{Chandrasekaran:2022cip}, that for quasi-dS space is of Type II$_\infty$ (where the trace of an operator diverges in the same limit) \cite{Seo:2022pqj, Gomez:2023wrq, Kudler-Flam:2024psh, Chen:2024rpx}.
This suggests that when the background  does not have a boundary, just like the dressing, the algebraic structure also seems to depend on whether the background breaks the isometry   or not.
Even though the mathematical equivalence between the dressed form of the relational observable and the von Neumann algebra is not fully clarified yet, it might be worth   investigating the implications of the former on the latter on the physical ground.
 To this end,  we focus on the Type II$_\infty$ structure of the operator algebra in quasi-dS space.

 In quasi-dS space, as in other backgrounds, the operators   are constructed from the smeared field observables.
  They are localized in some region of spacetime, for example, the region inside the horizon (which becomes the static patch in the $\epsilon_H \to 0$ limit).
 Then, a  state  in the Hilbert space can be constructed by acting on some normalizable state $|\Psi\rangle$ with an operator $a$ in the algebra, i.e., $a|\Psi\rangle$, which is  faithful provided the state is normalizable and its norm is positive semi-definite (with a vanishing norm occurring only for the zero operator). 
 For this purpose, the operator belonging to the algebra is given by  specific functional form  of the smeared field observable with bounded eigenvalues.
 \footnote{For example, we can consider the Weyl operator $W(f)=e^{i \varphi_f}$ where $\varphi_f$ is the smeared field observable with smearing function $f$.}
 Moreover, any state in the Hilbert space can be constructed in this way when $|\Psi\rangle$ is  `cyclic' (only the zero vector is orthogonal to all the states in the form of $a|\Psi\rangle$)  and `separating' ($a=0$ is the only operator satisfying $a|\Psi\rangle=0$).
 
 In the case of   dS space, the Bunch-Davies vacuum $|\Psi_{\rm BD}\rangle$ can be taken as a cyclic and separating state.
 This comes from the fact that $|\Psi_{\rm BD}\rangle$ is invariant under  dS isometries, which allows a function in the form of $\langle \chi |a |\Psi_{\rm BD}\rangle$ for any state $|\chi\rangle$ to be analytically continued to the entire complexified spacetime.
 However, in the absence of dS isometries, this argument can no longer be used to establish the existence of a cyclic and separating state.
 Even in this case, we may define a set of `adiabatic vacua'  $\{|\Psi_t\rangle\}$ which has been shown to exist in  FRW spacetime including both dS and quasi-dS space \cite{Parker:1969au, Fulling:1989nb, Lueders:1990np}.
 Here an adiabatic vacuum $ |\Psi_t\rangle $ can be understood as  the instantaneous ground state in the sense that no particles are created at the moment $t$.
 It is a kind of `Hadamard state', a state which looks like the Minkowski vacuum at sufficiently short distance \cite{Junker:1995wr}, implying that it is cyclic and separating with respect to the local algebra (for an overview, see, e.g., Section III of \cite{Kudler-Flam:2023qfl}).
 In the  dS limit where $\epsilon_H \to 0$, $|\Psi_{\rm BD}\rangle$ corresponds to the adiabatic vacuum which looks like the no-particle state at $t\to -\infty$.
 \footnote{  We note that $t$ here is time in the flat coordinates (see \eqref{eq:flatMet}), which must be distinguished from the static time $t_s$ in \eqref{eq:conv}.
 For dS space, we infer from the relation $\partial_{t_s}=\partial_t-Hr\partial_r$ that $H$, the Hamiltonian in the direction of $t$, and $H^{(s)}$, that in the direction of $t_s$, are related by $H^{(s)}=H-Hr P_r$ where $P_r$ is a momentum in the direction of $r$.
 Since the timelike dS  isometry is the translation in the direction of $t_s$, rather than $t$, $|\Psi_{\rm BD}\rangle$ is annihilated by $H^{(s)}$, not by $H$.
 Indeed, at $t\to -\infty$, $|\Psi_{\rm BD}\rangle$ behaves as a vacuum where particle is not created at all.
 As $t$ increases, $|\Psi_{\rm BD}\rangle$ becomes a multi-particle state, which is reflected in the Bogoliubov transformation between pairs of the creation and annihilation operators of particle at different values of $t$. \label{footnote:frame}}
  Then so far as $\epsilon_H$ is sufficiently small, we expect that the adiabatic vacuum $|\Psi_{-\infty}\rangle $ of quasi-dS space (the no-particle state at $t\to -\infty$) is very close to $|\Psi_{\rm BD}\rangle$ : it can be perturbatively expressed  as $|\Psi_{-\infty}\rangle =   |\Psi_{\rm BD}\rangle +\epsilon_H |\Psi^{(1)}\rangle+\epsilon_H^2 |\Psi^{(2)}\rangle+\cdots$, where $|\Psi^{(1)}\rangle$ and $|\Psi^{(2)}\rangle$ as well as  $|\Psi_{\rm BD}\rangle$ are dS states.
 When the function $\langle\chi |a|\Psi_{-\infty}\rangle$ (where $|\chi\rangle$ and $a$ are now a state and an operator in quasi-dS space which may be perturbatively expressed in terms of dS ones) is analytic  to sufficiently higher order of $\kappa$, which is typically assumed in previous discussions on von Neumann algebra \cite{Seo:2022pqj, Gomez:2023wrq, Kudler-Flam:2024psh, Chen:2024rpx} as well as cosmological perturbation theory, the cyclic and separating property of  $|\Psi_{\rm BD}\rangle$ also can be found in $|\Psi_{-\infty}\rangle$.
 Pairs of the creation and annihilation operators of particle at different values of $t$ are related by Bogoliubov transformation, and  adiabatic vacua (the no-particle states at  different values of $t$) have the same structure, which supports the local cyclic and separating property of the adiabatic vacuum.

 Now, we discuss the Type II$_\infty$ property of the von Neumann algebra for quasi-dS space.
 In order to observe the features of Type II$_\infty$ von Neumann algebra, we consider the behavior of the matter part of the Hamiltonian $H_M$.
For example, in order to describe  black hole as seen by an observer at asymptotic infinity, we need the matter Hamiltonian   given by $\int_\Sigma d\Sigma T_{\mu\nu}k^\mu n^\nu$ ($\Sigma$ : Cauchy surface of the region between the observer and the horizon,  $n^\nu$ : the unit normal to $\Sigma$, $k^\mu$ : timelike Killing vector), but the fluctuation  of it with respect to the Hartle-Hawking state $|\Psi_{\rm HH}\rangle$ diverges in the $\kappa \to 0$ limit.
To understand this divergence, we consider  the boundary dynamics, which encodes information on black hole in the bulk according to the holographic principle.
Time translation of the boundary is generated by the ADM Hamiltonian  and its expectation value with respect to $|\Psi_{\rm HH}\rangle$ is the ADM mass.
  It diverges in the $\kappa \to 0$ limit as can be noticed from the fact that the ADM mass for the Schwarzschild black hole is given by  $M_{\rm ADM}=r_s/2G=4\pi r_s/\kappa^2$ where $r_s$ is the Schwarzschild radius.
Then to obtain the   time translation generator which does not diverge with respect to $|\Psi_{\rm HH}\rangle$ in the $\kappa \to 0$ limit, we renormalize the matter   Hamiltonian as $H_{M, {\rm ren}}=H_M-\langle \Psi_{\rm HH}| H_M|\Psi_{\rm HH}\rangle=H_M-M_{\rm ADM}$, such that  $\langle \Psi_{\rm HH} | H_{M, {\rm ren}}|\Psi_{\rm HH}\rangle$ does not diverge but is fixed to $0$.
But still, the state $H_{M, {\rm ren}}|\Psi_{\rm HH}\rangle$ is not   well defined when its norm $\langle \Psi_{\rm HH}|(H_{M, {\rm ren}})^2|\Psi_{\rm HH}\rangle$ which is nothing more than the squared uncertainty (in other word, the fluctuation) of $H_M$  diverges in the  $\kappa \to 0$ limit.
Consequently, although $H_M$ itself is positive, the sign of the fluctuation is not restricted,    then we expect that the spectrum of $H_{M, {\rm ren}}$ ranges over $(-\infty, \infty)$ in the $\kappa \to 0$ limit.
%When we consider a two-sided eternal black hole, the Hamiltonian of the left patch and that of the right patch, or equivalently, the left and the right boundary Hamiltonians fluctuate as described above independently, then for an observer in one of the patches    
For details, see \cite{Witten:2021unn, Chandrasekaran:2022eqq}.

A similar situation arises  in quasi-dS space.
  To see this, we recall  that while $|\Psi_{-\infty}\rangle \simeq |\Psi_{\rm BD}\rangle$  is interpreted as an approximate vacuum (thus the energy eigenstate) of $H^{(s)}$ (the Hamiltonian generating $t_s$ translation, as we noted in  footnote \ref{footnote:frame}), it serves as  an approximate vacuum of $H$ (the Hamiltonian generating $t$ translation) only at $t\to -\infty$.
At a finite value of $t$,   $|\Psi_{-\infty}\rangle$ becomes a multi-particle state,  and as we will see below, the fluctuation of $H$ with respect to  $|\Psi_{-\infty}\rangle$ increases over time.
One may instead consider another adiabatic vacuum $|\Psi_{t_0}\rangle$ which acts as the  vacuum   of $H$ at $t_0$.
 But again, $|\Psi_{t_0}\rangle$ is not a vacuum of $H$ at  $t\ne t_0$, and the fluctuation of $H$ with respect to $|\Psi_{t_0}\rangle$ grows as $|t-t_0|$ increases.
Since we dressed the local operator by $H_M$ (the matter part of $H$ : we can treat $H_M$ and $H_G$ separately in the $\kappa\to 0$ limit where gravity decouples from matter), not by $H_M^{(s)}$ (the matter part of $H^{(s)}$) for gauge invariance, we need to observe the fluctuation of $H_M$ with respect to the adiabatic vacuum $|\Psi_t\rangle$:
\begin{itemize}
\item    $\langle \Psi_t| H_M (t) |\Psi_t\rangle$ can be obtained by calculating $\int_\Sigma d\Sigma T_{\mu\nu}k^\mu n^\nu$ ($\Sigma$ : Cauchy surface of $t=$constant in the local region,  $n^\nu$ : the unit normal to $\Sigma$, $k^\mu$ : time direction along which Hamiltonian is defined, i.e., $\partial_t$ in our case), where $T_{ab}=\nabla_a \phi \nabla_b \phi-(\frac12(\nabla \phi)^2+V)g_{ab}$.
In the flat coordinates, multiplying $T_{tt}\simeq ({3}/{\kappa^2})H^2$ in the $\epsilon_H \to 0$ limit   by the horizon volume   $({4\pi}/{3})H^{-3}$ (while the horizon radius in the flat coordinates is given by $(aH)^{-1}$, an additional factor $a^3$ comes from $\sqrt{-g}$) we obtain the expectation value of $H_M$ inside the horizon, $\langle \Psi_t| H_M(t)|\Psi_t\rangle = {4\pi}/({\kappa^2 H})$.
Moreover, while $H_M(t)$ is time-dependent, it varies very slowly because   $H$ is nearly constant.
Then we also expect that $\langle \Psi_t| H_M(t')|\Psi_t\rangle\simeq  {4\pi}/({\kappa^2 H})$ even for $t\ne t'$.
This expectation value of the Hamiltonian in fact coincides with $TS$, where $T= {H}/({2\pi})$ and $S= {A}/({4G})= {8\pi^2}/({\kappa^2 H^2})$ are the temperature and the entropy of the horizon, respectively, and  diverges as $ {1}/{\kappa^2}$ in the $\kappa\to 0$ limit. 
This suggests that as in the case of the black hole, $H_M$ in quasi-dS space needs to be renormalized as  $H_{M, {\rm ren}}(t)=H_M(t)-\langle \Psi_t | H_M(t)|\Psi_t\rangle$.
Indeed, the divergence of the expectation value of the Hamiltonian in the $\kappa \to 0$ limit can be found in other coordinate systems, i.e., for different choices of $k^\mu$.
For example, in \cite{Seo:2022pqj} (and as sketched in Appendix \ref{App:qdScase}), the energy flux across the horizon was considered.
Since approximate Killing vector $k^\mu=(\partial_t-Hr\partial_r)^\mu \simeq(\partial_{t_s})^\mu$ is tangent to the horizon, it yields the first law of thermodynamics $d\langle\Psi_{\rm BD}|H^{(s)}_{M}|\Psi_{\rm BD}\rangle \sim TdS $, implying that $\langle\Psi_{\rm BD}|H^{(s)}_{M}|\Psi_{\rm BD}\rangle$ diverges as $1/\kappa^2$ in the $\kappa \to 0$ limit (see also \cite{Frolov:2002va, GalvezGhersi:2011tx}).
\item  We may consider the renormalized Hamiltonian given by $H_{M, {\rm ren}}(t) = H_M (t) -\langle H_M(t)\rangle_{t_0}$, where $\langle H_M(t)\rangle_{t_0}=\langle \Psi_{t_0} |H_M(t)|\Psi_{t_0}\rangle$ is the vacuum expectation value of $H_M(t)$ with respect to some adiabatic vacuum $|\Psi_{t_0}\rangle$. 
While $\langle \Psi_{t_0} |H_M(t)|\Psi_{t_0}\rangle$ is almost constant in $t$, $|\Psi_{t_0}\rangle$ is an eigenstate of $H_M(t)$ only at $t=t_0$, which is reflected in the growth of the  fluctuation   $\delta H_M(t)$ as $\Delta t=|t-t_0|$ increases. 
Physically, it originates from the curvature perturbation modes stretched beyond the horizon.
They behave  as the classical fluctuation, providing the fluctuation of the clock $\phi_0$ (or $H$) hence that of time \cite{Vilenkin:1982wt, Linde:1982uu, Starobinsky:1982ee, Linde:1986fd}:
\dis{&(\delta \phi_0)^2=\Big(\frac{H}{2\pi}\Big)^2 (H\Delta t),
\\
&|\delta H|=|\dot{H}\delta t|=\Big|\frac{\dot H}{\dot \phi_0}\delta \phi_0\Big|=\frac{\sqrt{\epsilon_H}}{2\sqrt2 \pi}\kappa H^2(H\Delta t)^{1/2}.\label{eq:random}}
Then we obtain  
\dis{|\delta H_M| = \frac{4\pi}{\kappa^2H^2}\delta H\sim \frac{1}{\kappa}\big(\epsilon_H H\Delta t\big)^{1/2},\label{eq:randomH}}  
which diverges as $1/\kappa$ in the $\kappa\to 0$ limit unless $\epsilon_H H\Delta t$ is close to $0$.
\footnote{For this, one may require that $\epsilon_H$ is not too small.
Indeed, so far as $\epsilon_H$ is nonzero, $\epsilon_H H\Delta t$ increases in time, and it becomes  ${\cal O}(1)$ at the end of inflation  when $\Delta t\sim  {1}/{(\epsilon_H H)}$.  }
Since $(\delta H_M)^2=\langle \Psi_{t_0}|(H_{M, {\rm ren}}(t))^2|\Psi_{t_0}\rangle$ is   the norm of the state $H_{M, {\rm ren}}(t)|\Psi_{t_0}\rangle$, its divergence  implies that the state $H_{M, {\rm ren}}(t)|\Psi_{t_0}\rangle$ is not normalized in the $\kappa\to 0$ limit.
From this and the  divergence of $\langle H_M(t)\rangle_{t_0}$ in the same limit, one finds that $H_{M, {\rm ren}}(t)$ does not belong to the algebra describing the matter sector in the $\kappa\to 0$ limit.
That is, the translation of $t$ is the outer automorphism, and the value of   $H_{M, {\rm ren}}(t)$  ranges over $(-\infty, \infty)$ in the $\kappa \to 0$ limit. 
\end{itemize}
%From this, as being done in the case of   black hole, we can define   the renormalized Hamiltonian $H_{M, {\rm ren}}=H_M-\langle \Psi| H_M|\Psi\rangle$, such that   $\langle \Psi|H_{M, {\rm ren}}|\Psi\rangle$ does not diverge but is fixed to $0$. 
%But this is not enough for  $H_{M, {\rm ren}}$  to belong to the algebra generating the state by acting on $|\Psi\rangle$, since  n $\delta H_M \sim 1/{\kappa}$   diverges in the $\kappa\to 0$ limit.
%This can be resolved by considering the rescaled Hamiltonian  $ (\kappa H) H_{M, {\rm ren}}$  as an operator generating a state from $|\Psi\rangle$.
%But the thermodynamic distribution considers  exp$[-\beta H_{M, {\rm ren}}]$ where $\beta=1/{T}$, rather than   exp$[-\beta (\kappa H) H_{M, {\rm ren}}]$, and the value of   $H_{M, {\rm ren}}$  ranges over $(-\infty, \infty)$ in the $\kappa \to 0$ limit.

Before discussing how the properties of   $H_M$  above lead to the Type II$_\infty$ algebra of quasi-dS space, for a comparison,  we   briefly review the findings of \cite{Chandrasekaran:2022cip} which established that  the  von Neumann algebra of dS space is of Type II$_1$.  
The main obstacle  in implementing the algebra of dS space was the absence of a dS background state that could serve as a clock.  
Unlike quasi-dS space, the dS background respects the timelike isometry, which is reflected in $\dot{H}=0$ and  $\dot{\phi}_0=0$, so   the local relational observable like the curvature perturbation is not defined.
Since dS space does not have a boundary as well, the nonlocal gravitational dressing is not available. 
In order to resolve this difficulty and  construct the gauge invariant relational observable in dS space, \cite{Chandrasekaran:2022cip} suggested to introduce an observer carrying a clock.  
Then the observer's Hamiltonian $h_{\rm obs}$ shifts the time $t_{\rm obs}$ defined in relation to the observer (for example, through the gravitational Wilson line connecting a spacetime point to the observer \cite{Giddings:2025xym}), satisfying $[t_{\rm obs}, h_{\rm obs}]=i$.
This modifies  the Hamiltonian constraint for gauge invariance   as $H_{\rm ADM}+h_{\rm obs}=H_G+H_M+h_{\rm obs}=0$.
From this, 
one can promote the local operator $O(x)$ to the dressed one,
 \dis{e^{-i H_{\rm ADM}t_{\rm obs} } O(x)e^{i H_{\rm ADM}t_{\rm obs} } ,\label{eq:Witten}}
 which commutes with the total Hamiltonian $H_{\rm ADM}+h_{\rm obs}$.
 We also   note that   introduction of an observer is required in the background with isometries, such as dS space, to identify a subregion in an invariant way.
This is because points in any region can be defined in a gauge invariant way provided they are identified in relation to the clock and the rod, which are absent when the background has timelike and spacelike isometries, respectively (see also discussion in Section 4 of \cite{Chen:2024rpx}).
In quasi-dS space, since the timelike dS isometry is broken,  a clock is already intrinsic to the background,  thus we do not need to introduce an observer as a clock separately.
Of course, since  spatial homogeneity and isotropy are maintained in quasi-dS space, we   still  need to introduce an observer playing the role of a rod, such that the dressed operator in the form of  \eqref{eq:Witten} with $H_{\rm ADM}t_{\rm obs}$ replaced by $P_i x_{\rm obs}^i$ can be constructed (see discussion in Section 3.2 of \cite{Seo:2025tsw}). 
 From this, we can completely define the local region  in an invariant way. 
Even in this case, the time coordinate of an observer is already identified in relation to the background, indicating that the fluctuation of time  is generated not by $h_{\rm obs}$ but by $H_G$, which is reflected in \eqref{eq:DrIn}: we do not need to introduce a fluctuating time with respect to $h_{\rm obs}$ separately  (see also discussion in \cite{Klinger:2026tws}).

To compare \eqref{eq:Witten} with the dressing we  discussed in Section \ref{Sec:Str}, we note that the local but undressed operator  $O_M(x)$ in \eqref{eq:DrO} consists solely of the matter field  and does not contain the metric fluctuation.
For gauge invariance,   translation by $q$ in the dressing  is generated by the gravitational part of the Hamiltonian $H_G$, which indeed is a typical way of obtaining the Schr\"odinger equation : replacing $H_G$ by $-i d/{dq}$ gives $i {d}/{dq}=H_M$.
In contrast, $O(x)$ in  \eqref{eq:Witten} contains both the matter field and the metric fluctuation, which is reflected in the fact that  $H_{\rm ADM}$ instead of $H_M$ appears in the dressing term.
Since time $t_{\rm obs}$ is defined in relation to the observer, it is a conjugate of $h_{\rm obs}$ rather than $H_G$.
  While the eigenvalue of $h_{\rm obs}$ can fluctuate, by requiring the positivity of it, one can define the trace of operators generated by the dressed dS operators and $h_{\rm obs}$ as
  \dis{{\rm Tr}(\cdot) =\int_0^\infty dh_{\rm obs}\beta e^{-\beta h_{\rm obs}}\langle \Psi| \cdot |\Psi\rangle, \label{eq:Witten2}}
  where $|\Psi\rangle$ is chosen to be the Bunch-Davies vacuum.
  This is equivalent to $\langle \widehat{\Psi}|\cdot |\widehat{\Psi}\rangle$, 
 where $|\widehat{\Psi}\rangle=|\Psi\rangle \otimes \sqrt{\beta}\int d h_{\rm obs}e^{-\beta h_{\rm obs}/2}|h_{\rm obs}\rangle$.
  Here $|\widehat{\Psi}\rangle$ extends $|\Psi\rangle$ (which encodes entanglement between different regions of spacetime) by incorporating the distribution of $h_{\rm obs}$ acting on the square integrable function of $h_{\rm obs}$. 
   \footnote{We note that once the observer is included, the Bunch-Davies vacuum   is not a physical state, since it is annihilated by $H_{\rm ADM}$, rather than the new total Hamiltonian   $H_{\rm ADM}+h_{\rm obs}$.
  It is chosen as the cyclic and separating state.
 Although the factorized form of $|\widehat{\Psi}\rangle$ reflects the decoupling between $O(x)$ and $h_{\rm obs}$ in the non-dynamical gravity limit   (where time is described as a classical parameter and $h_{\rm obs}$ behaves as a center of the algebra), the dressing required by gauge invariance as well as the weight $e^{-\beta h_{\rm obs}}$ allow one to construct the appropriate thermodynamic description when gravity becomes dynamical (see also discussion below (2.13) of  \cite{Witten:2021unn}).  }
  For  the dressed operator \eqref{eq:Witten}, one can use $H_{\rm ADM} | \Psi\rangle=0 $ to show that
  \dis{{\rm Tr}(e^{-i H_{\rm ADM}t_{\rm obs} } O(x)e^{i H_{\rm ADM}t_{\rm obs} } )=\int_0^\infty dh_{\rm obs}\beta e^{-\beta h_{\rm obs}}\langle \Psi| O(x) |\Psi\rangle.}
 In the integration over $h_{\rm obs}$,  the upper bound on $h_{\rm obs}$ which was originally $ {1}/{\kappa}$ (the Planck scale) is given by infinity, meaning that  the limit $\kappa\to 0$ has been taken.
 Since Tr$(1)=1$, i.e., the trace of identity does not diverge but can be normalized to $1$, the algebra of the operator in dS space has been claimed to be the Type II$_1$ von Neumann algebra.

  Meanwhile, both dressings of the matter field operator $O_M(x)$ in \eqref{eq:Drflat} and \eqref{eq:DrIn} employ $H_G$ as the generator of time (more precisely, $q$ in \eqref{eq:DrO}) translation, following the typical construction of the Schr\"odinger equation.
  This is formulated by expressing the  time fluctuation  as a functional of the metric fluctuation  given by $V$ and $-({\kappa}/{H})\zeta$, respectively.
 Since the clock is intrinsic to the  geometry in these cases, it is tempting to define trace without introducing  an additional   observer.
  For this purpose, as we have seen in the case of quasi-dS space and black holes (see \cite{Witten:2021unn, Chandrasekaran:2022eqq}), we need to take into account the fact that the expectation value as well as the fluctuation of $H_M$ diverge  in the $\kappa \to 0$ limit.
  In the absence of a boundary, the  Hamiltonian constraint for gauge invariance reads $H_{\rm ADM}=H_G+H_M=0$ then we can define the renormalized gravity part of Hamiltonian as $H_{G, {\rm ren}} = H_G+\langle H_M\rangle=-H_{M, {\rm ren}}$ such that $H_{\rm ADM}=H_{G, {\rm ren}}+H_{M, {\rm ren}}=0$ is satisfied.  
  Then the divergences of $\langle H_M\rangle$ and $\delta H_M$ in the $\kappa\to 0$ limit imply corresponding divergences in $\langle H_G\rangle$ and $\delta H_G$.
Moreover, since  $H_G$ generates the translation of time fluctuation,   it plays the same role as $h_{\rm obs}$ in \eqref{eq:Witten2} : the only difference is that it ranges over $(-\infty, +\infty)$ in the  $\kappa \to 0$ limit, while $h_{\rm obs}$ is restricted to  $(0, +\infty)$.

  We now   define the trace of operators generated by the local relational observables in the matter sector (given by \eqref{eq:DrIn}) and (renormalized) $H_G$.
 For this purpose, we consider a type III algebra ${\cal A}$ describing quantum  fields in the $\kappa \to 0$ limit where the background is no longer dynamical.
 Then an element of ${\cal A}$ corresponds to $O_M(t)$, a localized (such that smeared in the spatial direction as $\int d^3x f( \mathbf{x}) O_M(t, \mathbf{x})$) and bounded operator made up of   matter fluctuation ($\varphi(x)$ in our case) only.
 Whereas time translation of $O_M(t)$ is generated by $H_M$,   since both $\delta H_{M, {\rm ren}}$ and $\langle \Psi_t| H_{M }(t)|\Psi_t\rangle$ diverge in the $\kappa \to 0$ limit, $H_M$ does not belong to ${\cal A}$.
 Consequently, the time translation is an outer automorphism.
 As $\kappa$ becomes finite,   $H_G$ (which needs to be rescaled as $(\kappa H)H_{G, {\rm ren}}$)  is no longer a central generator, and it acts on the space of square integrable function.
 As we have seen, since the value of $H_G$ ranges over $(-\infty, +\infty)$ in the $\kappa \to 0$ limit, the promoted algebra generated by 
 \footnote{  While we use $H_{M, {\rm ren}}$ instead of $H_{M}$ for $O_{\rm dr}$, this substitution does not introduce any changes, because both   $\langle \Psi_t| H_{M }(t)|\Psi_t\rangle  $ (a $c$-number) and the gravity sector operator $\zeta$ commute with $O(x)$.
 Moreover, it is natural to consider both $H_{M, {\rm ren}}(t)$ and $O(t)$ at the same instant   because the commutation relations of local operators are defined on a specific time slice, say, $t=t_0$.}
  \dis{O_{\rm dr}[\varphi (t)]=e^{-i H_{M, {\rm ren}}(t)[-\frac{\kappa}{H}\zeta]}O_M(t) e^{iH_{M, {\rm ren}}(t)[-\frac{\kappa}{H}\zeta]} \quad {\rm and}\quad H_G,}
    becomes Type II$_\infty$.
 Comparing this  with the algebra of dS space generated by
 \dis{e^{-i H_{\rm ADM}t_{\rm obs} } O(x)e^{i H_{\rm ADM}t_{\rm obs} } \quad {\rm and}\quad h_{\rm obs}, }
 one finds that two algebras share the same structure, except that the allowed values of   $h_{\rm obs}$ (which plays the same role as  $H_G$) ranges over $(0, +\infty)$, which leads to Type II$_1$ algebra.

 Now, consider $O_{\rm dr}$ at some instant $t=t_0$.
 Then as a cyclic and separating state, we can choose the adiabatic vacuum $|\Psi_{t_0}\rangle$ (a no-particle state at $t_0$ only), which annihilates $H_{M, {\rm ren}}(t_0)$. 
 For $t\ne t_0$, this state is no longer an eigenstate of $H_{M, {\rm ren}}(t)$ : $\delta H_M$ grows as $\Delta t=|t-t_0|$ increases and diverges in the $\kappa\to 0$ limit, which leads to the Type II$_\infty$ property of  algebra in quasi-dS space.
 Then one can define  trace of Type II$_\infty$ by $\langle \widehat{\Psi}_{t_0}|\cdot |\widehat{\Psi}_{t_0}\rangle$, where $|\widehat{\Psi}_{t_0}\rangle=|\Psi_{t_0}\rangle \otimes \sqrt{\beta}\int d h_G e^{-\beta h_G/2}|h_G\rangle$  with $H_G|h_G\rangle=h_G|h_G\rangle$ (see (3.39) of \cite{Witten:2021unn}).
  More explicitly, 
  \footnote{  In this case,  the modular operator for $|\Psi_{t_0}\rangle$ is given by  $\Delta_{\Psi_{t_0}}=e^{-\beta H_{M,{\rm ren}}(t_0)}$ such that  $\Delta_{\Psi_{t_0}}|\Psi_{t_0} \rangle=|\Psi_{t_0} \rangle$.
Then $|\Psi_{t_0}  \rangle$ satisfies the KMS condition, $\langle\Psi_{t_0} |a_u b  |\Psi_{t_0}  \rangle= \langle\Psi_{t_0}  |b a_{u+i}  |\Psi_{t_0}  \rangle$ where $a_u=\Delta_{\Psi_{t_0} }^{-iu} a \Delta_{\Psi_{t_0} }^{iu}$.
Moreover, we can also construct the modular operator for $|\widehat{\Psi}_{t_0}\rangle$ by defining the Fourier transform of the operator $O_M$ by $\int du O_M(u)e^{i u \beta (H_M+H_G)}$, which leads to Tr$(O_M)=2\pi \langle\Psi_{t_0}  |O_M(i) |\Psi_{t_0}  \rangle $.
Together with the KMS condition, this is used to show the basic property of the trace,
\dis{{\rm Tr}\big(O_{M, 1}O_{M, 2}\big)=2\pi \int du \langle\Psi_{t_0}  |O_{M,1}(u) (O_{M, 2})_u (i-u) |\Psi_{t_0}  \rangle ={\rm Tr}\big(O_{M, 2}O_{M, 1}\big).}
For details, we refer the reader to Section 3.3 and 3.4 of  \cite{Witten:2021unn}.
 } 
   \dis{{\rm Tr}(\cdot) = \beta\int_{-\infty}^\infty d h_{G} e^{ - \beta h_{G}}\langle \Psi_{t_0} | \cdot |\Psi_{t_0} \rangle. \label{eq:Tr} } 
  Then the trace of the operators in the form of \eqref{eq:DrIn} is given by
\dis{{\rm Tr}\big(O_{\rm dr}(t_0)\big)&={\rm Tr}\big( e^{-iH_M(t_0)[-\frac{\kappa}{H}\zeta]}O_M[\varphi(t_0)]e^{i H_M(t_0)[-\frac{\kappa}{H}\zeta]}\big)
   \\
   &=\Big[\beta\int_{-\infty}^\infty d h_{G} e^{- \beta h_{G}}\Big]\langle \Psi_{t_0} | O_M[\varphi]  |\Psi_{t_0} \rangle.}  
   Factorizing the divergent integral inside the bracket yields $\langle \Psi_{t_0} | O_M[\varphi(t_0)]  |\Psi_{t_0} \rangle$, which is interpreted that $|\Psi_{t_0}\rangle\langle\Psi_{t_0} |$ gives the same effect as the density matrix $\rho=1$ of the maximal mixing \cite{Witten:2018zxz}. 
   This in fact has been one of motivations to investigate thermodynamic behavior of quantum gravity in the language of the von Neumann algebra.
    The divergent factor in Tr$(O_{\rm dr})$, which is also reflected in Tr$(1)=\infty$, i.e., the trace of the identity cannot be normalized to $1$, shows that the algebra of quasi-dS is given by the Type II$_\infty$ von Neumann algebra.
Meanwhile, for any functional $G(H_G)$, 
\dis{{\rm Tr}\big(G(H_G)\big)=\beta\int_{-\infty}^\infty d h_{G} e^{- \beta h_{G}}G(h_G)} 
  is satisfied, which is obviously interpreted as the thermal average of $G( H_{G})$.

     We can compare this with the findings of \cite{Kudler-Flam:2024psh, Chen:2024rpx}, which also concluded that the algebra of quasi-dS is of Type II$_\infty$.
    In order to describe the stochastic time evolution of the classical inflaton field, \cite{Kudler-Flam:2024psh, Chen:2024rpx} solved the Fokker-Planck equation.
   Here the   classical inflaton field exhibits   stochastic behavior driven by the fluctuations stretched beyond the horizon (which lose quantum coherence through decoherence and effectively behave  as the classical fluctuation) \cite{Burgess:2014eoa}.
   This classical fluctuation is nothing more than that of the scalar field (`clock') $\delta \phi_0$ in \eqref{eq:random}.
   Analysis in \cite{Kudler-Flam:2024psh, Chen:2024rpx} shows that the probability distribution contains a factor
   \dis{{\rm exp}\Big[-\frac{8\pi^2}{3 H^4}\frac{dV}{d\phi}\delta \phi_0\Big].}
   Meanwhile, since the background is close to dS space, the Friedman equation gives $H^2\simeq(\kappa^2/3)V$, the time derivative of which  leads to $\epsilon_H\simeq [1/(2\kappa^2)] V^{-2}(dV/d\phi)^2$ (see  \eqref{eq:epsilon}).
  Since the inverse temperature $\beta$ is  given by $2\pi/H$, the above factor can be rewritten as 
  \dis{{\rm exp}\Big[-\beta \frac{4\pi \sqrt{2\epsilon_H}}{\kappa H} \delta \phi_0\Big]\sim e^{-\beta \delta H_M},}
where for the last relation we refer to \eqref{eq:randomH} as well as \eqref{eq:random}.
Since $\delta \phi_0 \in (-\infty, +\infty)$, we again conclude that the probability distribution contributed by the fluctuation of time results in Type II$_\infty$ algebra of quasi-dS space.
On the other hand, in our approach, the connection between $\delta \phi_0$ and  the time fluctuation becomes evident through the dressing required by gauge invariance.

 We also note that by using  $H_{\rm ADM}$ instead of $H_M$ to define the dressing \eqref{eq:Witten} in dS space, the gravity sector consisting of the graviton excitations is treated as dynamical  even in the  $\kappa \to 0$ limit.
 This is quite natural because   the graviton still remains as the low energy degrees of freedom in this limit.
 Instead,   the interaction between the graviton and  matter is turned off, such that the gravity and the matter sectors form the independent Hilbert spaces which do not interact with each other.
 The similar situation can be found in quasi-dS space as well.
 As in the case of dS space, the graviton, the transverse traceless fluctuation of the metric, in quasi-dS space decouples in the $\kappa\to 0$ limit.
Moreover, given  the curvature perturbation with the canonical kinetic term $-\sqrt{2\epsilon_H}{\cal R}=\varphi-\sqrt{2\epsilon_H}\zeta$, $\zeta$ from the gravity sector and $\varphi$ from the matter sector decouple in the $\epsilon_H\to 0$ limit \cite{Green:2024hbw}.
Typically, this corresponds to the dS limit: since timelike isometry is restored, there is no reason for $\varphi$ to mix with $\zeta$.
On the other hand, we infer from \eqref{eq:epsilon} that the vanishing $\epsilon_H$ can be attributed to taking the $\kappa\to 0$ limit, while $\dot{\phi}_0$ is kept finite.
More precisely, one may require that even in the $\epsilon_H\to 0$ limit, the ratio   $\epsilon_H S=4\pi^2\dot{\phi}_0^2/H^4  $   which is independent of $\kappa$ can be sizeable.
   Since the condition $\epsilon_H S>1$ is equivalent to  $\epsilon_H>\kappa^2H^2$,  \footnote{Intriguingly, as pointed out in \cite{Seo:2022pqj}, this condition  is known to be the condition that the fluctuation of $\phi_0(t)$ generated by  the modes stretched beyond the horizon is small enough to forbid eternal inflation. For recent discussions, see, e.g., \cite{Kinney:2018kew, Brahma:2019iyy, Rudelius:2019cfh, Wang:2019eym, Seo:2022uaz}. \label{foot:int}}
   the factor $\sqrt{2\epsilon_H}$ multiplying $\zeta$ in the combination $\varphi-\sqrt{2\epsilon_H}\zeta$ must be larger than ${\cal O}(\kappa H)$.
 Consequently, even if $\epsilon_H$ is small, an ${\cal O}(\sqrt{\epsilon_H})$ contribution to a physical quantity relevant to mixing between $\varphi$ and $\zeta$ is not suppressed compared to an ${\cal O}(\kappa)$ contribution in the same quantity.   
 In this case,   the gravity and the matter sectors can   approximately decouple, preserving the features of the isometry breaking by the background.
 In any case, in the decoupling limit $\kappa\to 0$ and $\epsilon_H\to 0$ (while $\epsilon_H S$ is kept sizeable), local fields in the gravity and the matter sectors can be dressed separately as \eqref{eq:Drh} and \eqref{eq:DrIn}, respectively.
Then the trace of operators generated by the local relational observables in the form of \eqref{eq:Drh} and (renormalized and rescaled) $H_M$  may be written by exchanging the roles of the matter and the gravity sectors from \eqref{eq:Tr} : 
  \dis{{\rm Tr}(\cdot) = \beta\int_{-\infty}^\infty d h_M e^{- \beta h_M}\langle \Psi| \cdot |\Psi\rangle, }
  where $h_M$ is the eigenvalue of $H_{M, {\rm ren}}$.
  This also diverges since the value of $H_{M, {\rm ren}}$   ranges over $(-\infty, \infty)$ in the $\kappa\to 0$ limit.
Therefore,  in quasi-dS space, we can consider the   matter sector   and the gravity sector  as independent Hilbert spaces of the low energy EFT, with the structure given by the Type II$_\infty$ von Neumann algebra.
\footnote{ The situation where both the matter and the gravity sectors  are relevant to the low energy EFT may suggest a following interesting possibility.
Motivated by the holographic description of two-sided AdS black hole \cite{Maldacena:2001kr}, we can understand the thermodynamic behavior of black hole  by constructing the thermofield double state $|{\rm TFD}\rangle=\sum_n c_n |E_n\rangle_R |E_n\rangle_L$, where the states in left and right wedges $|E_n\rangle_{R/L}$ are entangled (see Section 2 of \cite{Witten:2021unn}).
This is annihilated by the `total boundary Hamiltonian' $H_{M, {\rm tot}}=H_{M,(R)}-H_{M,(L)}=H_{M,{\rm ren},(R)}-H_{M, {\rm ren}(L)}$  (the relative minus sign reflects the opposite directions of time in two wedges)   and $(H_{M, {\rm tot}})^2$.
Presumably, even if the background  has only one of wedges, say, the right wedge, and the background does not have a  boundary,  we may consider the state similar to the thermofield double by replacing roles of $H_{M, (R)}$ and $H_{M, (L)}$ by $H_M$ and $-H_G$, respectively since both $H_{\rm ADM}=H_M+H_G$   and $(H_{\rm ADM})^2$ must annihilate the physical state. 
Then  $|E_n\rangle_L$ would be the states in the gravity sector.}

 \section{Conclusion} 
 
  As we have seen, the relational observable achieving gauge invariance can be expressed as the dressed operator.
  When the background has a boundary where the diffeomorphism is not gauged, we can construct the nonlocal gravitationally dressed operator using the gravitational Wilson line. 
  In contrast, the isometry breaking background admits the local dressing, as certain fluctuations in the gravity and the matter sectors shift under the diffeomorphism in the direction of the broken isometry.
  Since the dressing takes the same form as the outer automorphism of the local operator in the von Neumann algebra, we can investigate the algebraic structure of the background through the dressing of relational observables.
  From this, we understand that quasi-dS background is described by the Type II$_\infty$ von Neumann algebra where the trace diverges in the $\kappa\to 0$ limit, whereas the von Neumann algebra of the dS background is of Type II$_1$ where the trace of the finite value is well defined in the same limit.
  This shows that the isometry preserving and breaking backgrounds are quite different in their algebraic structure, even though the breaking effect is small.
   We also find that not just the matter sector, the gravity sector in the low energy EFT can be accommodated in the algebra constrained by gauge invariance as both can be expressed in the dressed form.

%

%\newpage

\appendix

%\section{Review on Klebanov-Strassler throat}
%\label{App:KSreview}

%\subsection{Geometry near the tip ($\eta \ll 1$ limit)}

\renewcommand{\theequation}{\Alph{section}.\arabic{equation}}

%\section{Uncertainty for the infrared modes}
%\label{app:IRuncert}
%\setcounter{equation}{0}

\subsection*{Acknowledgements} 

This work is motivated by an anonymous referee's comment on \cite{Seo:2025tsw}.

 \appendix
 \section{Hamiltonian in quasi-dS space}
 \label{App:qdScase}

  In this appendix, we provide  a brief review on divergences of $\langle \Psi| H^{(s)}_M|\Psi\rangle$ and  $\langle \Psi|(H_{M, {\rm ren}})^2|\Psi\rangle$ in the $\kappa \to 0$ limit, which was   discussed in \cite{Seo:2022pqj}.
To begin with, consider  $\langle \Psi| H^{(s)}_M|\Psi\rangle$,  where $|\Psi\rangle$ is  given by the spacetime state which converges to the Bunch-Davies vacuum in the $\epsilon_H \to 0$ limit.
  The variation of $\langle \Psi| H^{(s)}_M|\Psi\rangle$ in time generated by the timelike dS isometry breaking is given by  the energy   flux across the boundary. 
While the boundary of black hole is given by the asymptotic infinity, we take the horizon as the boundary of quasi-dS space.
For the  dS background, there exists a timelike isometry associated with the timelike Killing vector $k^a=(\partial_{t_s})^a=(\partial_t-Hr\partial_r)^a$.
 In this case, one may consider the conserved energy  $H^{(s)}_M=\int_\Sigma d\Sigma T_{\mu\nu}k^\mu n^\nu$ by taking the horizon as a boundary 
 ($\Sigma$ : a Cauchy surface of the region between the static observer's worldline and the horizon, $n^\nu$ : the unit normal to $\Sigma$).
 This timelike isometry is broken in the quasi-dS background where $k^a$ is admitted as  the approximate timelike Killing vector  only when $\epsilon_H$ is very tiny.
 Then the energy written in the same expression as that in dS space is no longer conserved, which is reflected in the nonzero energy flux across the horizon.   
 To see this, we note that the metric in the flat coordinates is given by \eqref{eq:flatMet} and the  horizon radius is $r=e^{-Ht}/H$ (that is, $r_S= {1}/{H}$ in the static coordinates : see \eqref{eq:conv}).
 Since $n^\mu=k^\mu$ on the future horizon ${\cal H}^+$, we obtain
\dis{T_{\mu\nu}k^\mu n^\nu\Big|_{\cal H^+}&=T_{\mu\nu}k^\mu k^\nu\Big|_{\cal H^+}=\Big[\Big(\frac12\dot{\phi}_0^2+V\Big)+(Hr)^2 e^{2Ht}\Big(\frac12\dot{\phi}_0^2-V\Big)\Big]_{\cal H^+}
\\
&=\dot{\phi}_0^2,}  
from which the energy flux across the horizon is given by 
 \dis{\delta H^{(s)}_M=\delta \int_{\cal H^+} d\Sigma T_{\mu\nu}k^\mu n^\nu=\frac{4\pi}{H^2} \delta t_s \dot{\phi}_0^2=\frac{8\pi}{\kappa^2}\epsilon_H \delta t_s, }
 where we use the fact that $t_s$ can be used as  the affine parameter along the null generator of the horizon. 
 Noting that $\partial t/\partial t_s=1$, this can be compared with the variation of the entropy
 \dis{\delta S&=\frac{\delta A}{4G}=-\frac{2\pi}{G}\frac{\dot H}{H^3}\delta t=\Big(\frac{2\pi}{H}\Big)\frac{8\pi}{\kappa^2}\epsilon_H \delta t
 \\
 &=\frac{1}{T}\frac{8\pi}{\kappa^2}\epsilon_H \delta t}
 to give $\delta H^{(s)}_M=T\delta S$, or $H^{(s)}_M \sim TS$ on the dimensional ground   \cite{Frolov:2002va} (see also  \cite{GalvezGhersi:2011tx, Seo:2022pqj}).
 We also find that $\delta H^{(s)}_M$ vanishes in dS space where $\epsilon_H =0$, as expected.
  For the finite value of $\epsilon_H$,  $\delta H^{(s)}_M$ hence the typical scale of $H^{(s)}_M$ diverges as $ {1}/{\kappa^2} $ in the   $\kappa \to 0$ limit, implying that the renormalization of  $H^{(s)}_M$ as  $H^{(s)}_{M, {\rm ren}}=H^{(s)}_M-\langle \Psi|H^{(s)}_M|\Psi \rangle$ is required.

 We now discuss the fluctuation of $H_M$.
We will find that  $H_{M, {\rm ren}}$ does not belong to the algebra of the low energy EFT,   since $\langle \Psi|(H_{M, {\rm ren}})^2|\Psi\rangle$,   the   norm of $H_{M, {\rm ren}}|\Psi\rangle$, diverges  in the $\kappa\to 0$ limit.
  To see this, we note that    $\langle \Psi|(H_{M, {\rm ren}})^2|\Psi\rangle$ is interpreted as $(\delta H_M)^2$, the squared fluctuation of $H_M$.
In the presence of the horizon, the curvature perturbation mode with the wavenumber $k$ is `frozen' when its wavelength is stretched beyond the horizon, i.e., at $t=H^{-1}\log\big( {k}/{H}\big)$ such that it behaves as the classical fluctuation. 
As time goes on, more and more modes are frozen in this way, contributing to the fluctuation of the classical trajectory of $\dot{\phi}_0$ to give \cite{Vilenkin:1982wt, Linde:1982uu, Starobinsky:1982ee, Linde:1986fd}
\dis{(\delta \phi_0)^2=\Big(\frac{H}{2\pi}\Big)^2 (H\Delta t).}
This shows that the fluctuation is proportional to the temperature and accumulated in time as $t^{1/2}$. 
Since the classical solution ${\phi}_0(t)$ plays a role of the clock, $\delta \phi_0$ is interpreted as the fluctuation of time, which generates  the fluctuation of another clock $H$ as 
\dis{|\delta H|=|\dot{H}\delta t|=\Big|\frac{\dot H}{\dot \phi_0}\delta \phi_0\Big|=\frac{\sqrt{\epsilon_H}}{2\sqrt2 \pi}\kappa H^2(H\Delta t)^{1/2}\sim \kappa H^2,}
where for the last estimation, we take $\Delta t$ to be the time scale of the inflationary period, $\Delta t \sim  {1}/({\epsilon_H H})$.
From this, we obtain
\dis{|\delta H_M| = \frac{4\pi}{\kappa^2H^2}\delta H\sim \frac{1}{\kappa}.\label{eq:HMfl}}

\end{document}